\documentclass[letterpaper,12pt]{article}
\usepackage{setspace}
\usepackage{tabu}
\usepackage{times}
\usepackage{booktabs}
\usepackage[table,xcdraw]{xcolor}
\usepackage{charter}
\usepackage[english]{babel}
\usepackage{wrapfig}
\usepackage{upgreek}
\usepackage{siunitx}
\usepackage[font=small,labelfont=bf]{caption}
\usepackage{titlesec}
\usepackage{graphicx}
\usepackage{pgfplots}
\usepackage{fancyhdr} 

\titleformat{\section}
  {\normalfont\fontsize{11}{12}\bfseries}{\thesection}{1em}{}
\titleformat{\subsection}
  {\normalfont\fontsize{11}{12}\bfseries}{\thesubsection}{1em}{}

\titlespacing*\section{0pt}{3mm plus 1pt minus 2pt}{1mm plus 2pt minus 2pt}
\titlespacing*\subsection{0pt}{3mm plus 1pt minus 2pt}{1mm plus 2pt minus 2pt}
\titlespacing*\subsubsection{0pt}{3pt plus 3pt minus 2pt}{0pt plus 2pt minus 2pt}
\titlespacing*\paragraph{0pt}{6pt plus 3pt minus 2pt}{0pt plus 2pt minus 2pt}

\graphicspath{ {./Figures/} }

\usepackage[letterpaper,top=1in,bottom=1in,left=1in,right=1in]{geometry}

\usepackage{amsmath}
\usepackage{graphicx}
\usepackage[colorinlistoftodos]{todonotes}
\usepackage{siunitx}
\usepackage{latexsym}
\DeclareSIUnit{\sqrthz}{\ensuremath{\sqrt{\text{\hertz}}}}
\pgfplotsset{compat = 1.14}

\title{Single-Photon Single-Flux Coupled Detectors}
\author{Murat Onen$^{1,2}$ \and Marco Turchetti$^{1,2}$, \and Brenden A. Butters$^{1,2}$, \and Mina R. Bionta$^{2}$, \and Phillip D. Keathley$^{2}$, \and Karl K. Berggren$^{1,2}$}

\begin{document}

%
%
%
%
%

\newpage
	
\date{\small $^1$ Department of Electrical Engineering and Computer Science, Massachusetts Institute of Technology, Cambridge, Massachusetts, 02139, USA\\ $^2$ Research Laboratory of Electronics, Massachusetts Institute of Technology, Cambridge, Massachusetts 02139, USA\\}
\maketitle

\setcounter{page}{1}
\noindent\textbf{Abstract:} 

In this work, we present a novel device that is a combination of a superconducting nanowire single-photon detector and a superconducting multi-level memory. We show that these devices can be used to count the number of detections through single-photon to single-flux conversion. Electrical characterization of the memory properties demonstrates single-flux quantum (SFQ) separated states. Optical measurements using attenuated laser pulses with different mean photon number, pulse energies and repetition rates are shown to differentiate single-photon detection from other possible phenomena, such as multi-photon detection and thermal activation. Finally, different geometries and material stacks to improve device performance, as well as arraying methods are discussed.

\noindent\textbf{Keywords: Single-photon detectors, single-flux electronics, superconducting devices} 

\clearpage
\setcounter{page}{1}
\pagestyle{fancy}
\section{Introduction}

Detection of single-photons is essential to applications such as optical communications, astronomical measurements, and quantum-information processing. Superconducting nanowire single-photon detectors (SNSPDs) are prominent tools for these applications because of their $>90\%$ detection efficiencies, near GHz count rates, few picosecond timing jitter, broad spectral sensitivity from ultra-violet to infrared wavelengths, and sub-Hz dark count rates \cite{natarajan2012superconducting, zhu2018scalable,zhao2017single}. 

SNSPDs operate by amplifying photon-absorption events through the growth of a hotspot (i.e. a normal domain in the superconductor), across which a significant (few $\SI{}{\milli\volt}$) voltage develops. However, the high kinetic inductance of the superconductor leads to slow recovery (i.e. longer period for hotspot to grow), causing the possibility of latching and increased power dissipation. If the hotspot's growth could be regulated, these effects could be avoided with the limiting case being the absorption of a single photon resulting in the passage of a single flux across the nanowire. 

In this paper, we attempt to achieve the minimal state of single-photon to single-flux conversion by shunting a nanowire both electrically and thermally with an underlying normal metal film. Unfortunately, direct measurement of the dynamics of single-flux events in a shunted superconductor is extremely challenging due to the low signal-to-noise ratio. Instead, we used a nanoSQUID to detect the fluxon-tunneling event post-facto. This approach has the attractive feature of storing fluxon state of the nanoSQUID as a persistent current in the loop. As a result, the number of firing events of the nanowire can be integrated and read out later. 

We suggest this new device may prove analogous to a charge-coupled detector (CCD), as the photon detection events lead to accumulation of flux (instead of charge for CCDs), which can provide the number of single-photon detection events (analogous to intensity). The further development could devise schemes for readout of large-scale arrays of such devices for use in imaging and spectroscopy.

\section{Device Operation}

\begin{figure}[h!]
	\centering
	\includegraphics[width = 0.7\linewidth]{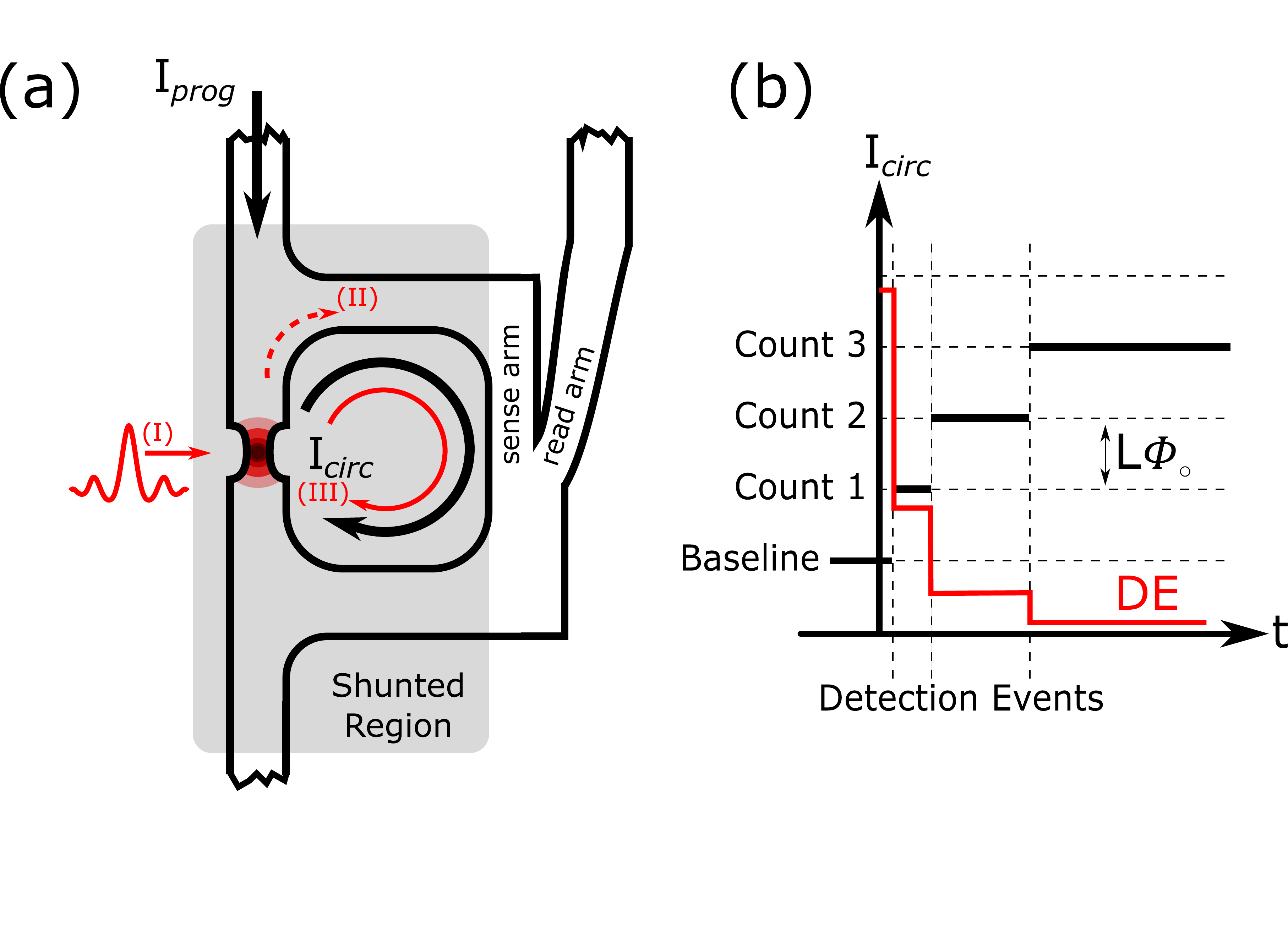}
	\caption{\textit{\textbf{Device schematic and operation:}} \textbf{(a)} Schematic representation of the device layout consisting of the shunted constriction (left) and the yTron (right). The programming current $I_\textrm{prog}$ is constantly provided to the device to bias the nanowire. Each photon registry switches the constriction into resistive (normal) state (I) and redirects the current from the constriction branch into the loop (II). This redirected current then gets trapped in the loop, once the nanowire restores to its superconducting state (III) \textbf{(b)} Notional graph of circulating current and the resultant detection efficiency versus time. The circulating current inside the unit cell is used to determine number of single-photon detection events. Increasing $I_\textrm{circ}$ reduces the effective bias ($I_\textrm{prog} -I_\textrm{circ}$) and therefore the detection efficiency.}
	\label{fig:fig1}
\end{figure}

The single-photon single-flux coupled detector demonstrated in this work is a superconducting loop that employs a shunted constriction \cite{toomey2019bridging} as the programming element (i.e. the active area of the detector) and a y-shaped current combiner (yTron) \cite{mccaughan2016} as the reading element (Figure \ref{fig:fig1}a). The device operates as a multi-level memory, where the state of the cell is stored by the persistent current that circulates in the loop. The state of the cell can be electrically programmed by applying a current signal that exceeds the switching current of the shunted constriction ($I_\textrm{SW}$), causing it to switch into a resistive (normal) state\footnote{Current flowing through the yTron branch is ignored based on the high inductance ratio, for simplicity of the analysis.}. This event redirects the current from the constriction branch into the loop. Under the absence of current flow, the nanowire restores its superconducting state and traps the current shuttled into the loop ($I_\textrm{circ}$). The shunting of the constriction limits the amount of current shuttling to a single-flux quantum (SFQ) by dampening the process electrically and thermally. A more detailed analysis of the operation can be found in Ref.\cite{onen2019design}.

The state of the cell can be read by the yTron terminal, using a geometric effect called current crowding \cite{mccaughan2016,clem2011geometry}. This effect manifests itself as a modulation on the switching current of the yTron’s read arm by the circulating current flowing in it sense arm. Therefore, the current at which the read arm switches can be used to measure the state of the cell nondestructively. As the device is non-volatile, the state readout can be performed multiple times at low frequencies.

The main idea behind this device is to trigger these switching events via optical illumination, as photons can create hotspots on superconducting nanowires biased closed to $I_\textrm{SW}$. Here, the constriction is more than 4 times narrower than the loop and the read arm. Therefore, when the constriction is critically biased, currents in other parts of the wire are well below their switching current, ensuring the photon-triggered counting events being predominantly from the constriction.

In this scenario, each detection event shuttles additional flux inside the loop, changing the state of the cell memory incrementally. Consequently, the state difference between the two successive readouts ($\propto \Delta I_\textrm{circ}$) gives the number of photons registered by the device in that time interval. It should be noted that each photon detection increases the circulating current in the loop, leading to a decrease in the effective bias over the shunted nanowire (Fig.\ref{fig:fig1}b). This reduced bias causes a super-linear decrease in detection efficiency \cite{marsili2013detecting}, similar to that of CCDs with charge accumulation \cite{janesick2001scientific}.

\section{Experiment}
\label{sec:method}

\begin{figure}[h!]
	\centering
	\includegraphics[width = \linewidth]{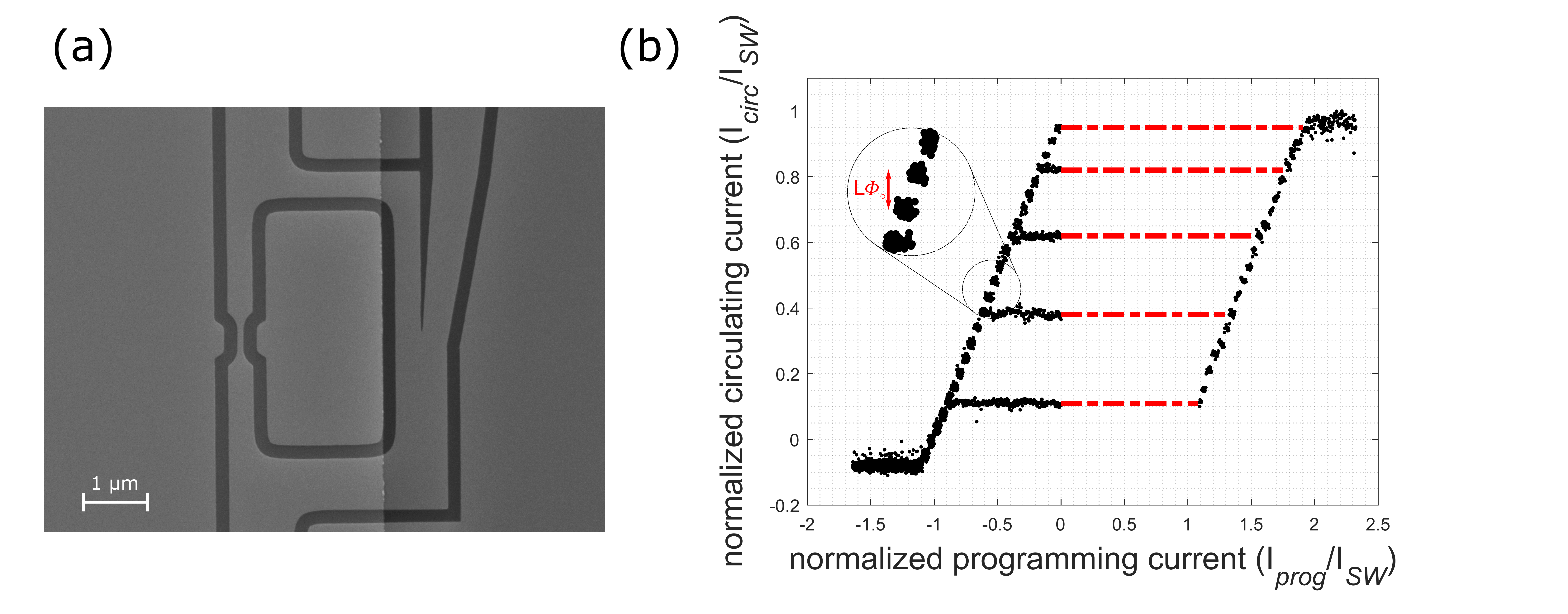}
	\caption{\textit{\textbf{Fabricated device and electrical characterizations:}} \textbf{(a)} Scanning electron micrograph of the device. Contrast difference in the middle indicates the edge of the metal shunting layer. \textbf{(b)} Electrical characterization of the device. The device showed 25 states which were separated by a single-flux quantum difference ($L\Phi_\circ$). Multiple traces are used to determine the circulating current,  $I_\textrm{circ}$ and programming current $I_\textrm{prog}$ normalized to the switching current $I_\textrm{SW}$ of the device at different states (See \ref{sec:electrical}). These results also show that at any given state, the effective bias is equal to $~0.95 I_\textrm{SW}$, which is comparable to the SNSPD bias levels for single-photon detection.}
	\label{fig:fig2}
\end{figure}

The devices described above were fabricated with \SI{25}{\nano\meter} thin-film NbN on a metal shunting stack of \SI{20}{\nano\meter} Ti capped with \SI{5}{\nano\meter}  Pt\footnote{The effective thickness of the superconducting layer is likely thinner \SI{10}{\nano \meter} due to the proximity effect. We have previously observed the stack was not superconducting at all at \SI{4.2}{\kelvin} when the NbN thickness was $\leq\SI{10}{\nano\meter}$. Therefore, we increased the NbN thickness to \SI{25}{\nano \meter}, a level that is unusual for single-photon sensing operations.}. Sheet resistance and critical temperature ($T_\textrm{C}$) of the unshunted superconducting NbN film were \SI{94}{\ohm}$/\Box$ and \SI{10.5}{\kelvin} respectively. The superconducting layer was patterned such that the constriction portion of the loop resides on the metal layer (in-situ shunting) and yTron side does not.  The constriction is \SI{100}{\nano\meter} wide while both yTron arms are \SI{400}{\nano\meter} wide with a connection angle of \SI{75}{\milli \radian}. Figure \ref{fig:fig2}a shows the scanning electron microgram of the fabricated structure.

We first characterized the device electrically in liquid helium (LHe) immersion conditions (\SI{4.2}{\kelvin}), by programming the device into different states to quantify the circulating current inside the device at each step as a fraction of the switching current of the shunted nanowire (Fig.\ref{fig:fig2}b). Considering that the device consists of 32.6 squares and an estimated kinetic inductivity ($L_\text{k}$) of \SI{12.4}{\pico\henry}$/\Box$, a single-flux in this loop should be equal to \SI{5.1}{\micro\ampere}. On the other hand, another stand-alone nanowire on this chip has a switching current $(I_\textrm{SW})$ of \SI{62}{\micro \ampere}, indicating that the separation of states is \SI{2.7}{\micro \ampere}, equivalent to $0.53\Phi_\circ$ (See Sec.\ref{sec:electrical}). Similar to discussions made in Ref.\cite{onen2019design}, we have attributed this discrepancy to underestimating the kinetic inductivity increase in the shunted region due to suppression of superconductivity (the calculations assume a single $L_\textrm{k}$ value which is estimated for the unshunted region). Furthermore, we have also observed that the relative alignment of the metal and the NbN heavily influenced the biasing conditions\footnote{The inductance ratio of the two arms is a strong function of what portion of the yTron branch is shunted. It can be seen from Fig.\ref{fig:fig2} that the optimal ratio was obtained when the metal layer was in contact with the yTron branch until the active readout area. Shunting of the yTron itself reduces the readout performance \cite{onen2019design}}. This behavior also suggests large kinetic inductivity difference between the shunted and the unshunted regions and requires placement of the metal layer as close as possible to (but not in contact with) the yTron.

\begin{figure}[h!]
	\centering
	\includegraphics[width = \linewidth]{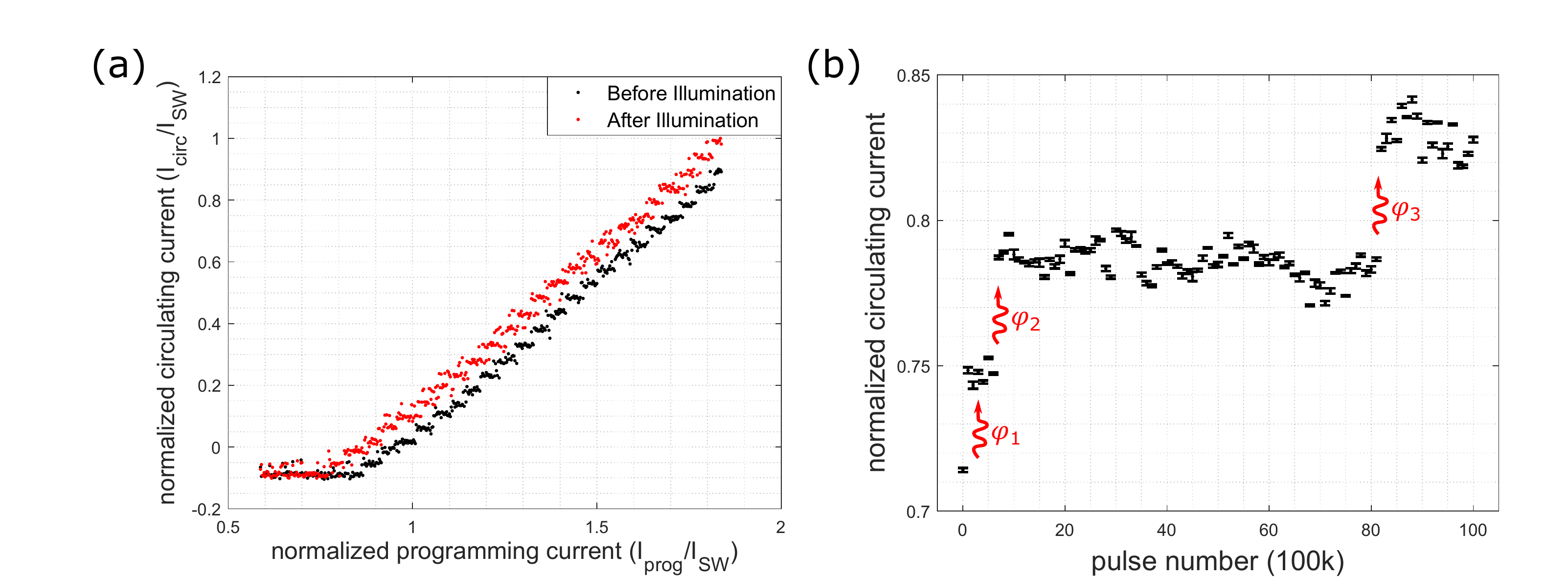}
	\caption{\textit{\textbf{Optical measurement of the device and single-photon-counting demonstration:}} \textbf{(a)}Optical modulation of the $I_\textrm{circ}$ for different initial programming conditions. The device is first set to a state (black trace) and then illuminated with a \SI{80}{\micro\watt} continuous waveform laser at \SI{680}{\nano \meter} for \SI{5}{\second} while keeping the same programming current. The state change of the device (i.e. the normalized circulating current difference between successive reads before and after the illumination) was found to be indifferent to the initial programming as expected, which can be seen as perfectly shifted state curves for before and after illumination readouts.\textbf{(b)} The device is initially set to an electrical state and then illuminated with an ultrafast laser centered at \SI{1030}{\nano \meter}, with a repetition rate of \SI{1}{\mega \hertz} and pulse energy  $\approx$\SI{80}{\pico \joule}.  A readout is performed following each pulse train to probe if the state has changed in that cycle. The state of the device changed by a single-flux quantum 3 times as a response to optical pulses. Note that the beam spot size is $\approx\SI{1}{\centi\meter\squared}$ for both cases.}
	\label{fig:fig3_and_4}
\end{figure}

For the optical characterization, we first used an \SI{80}{\micro \watt}, continuous waveform laser centered at \SI{680}{\nano \meter} to illuminate the sample at different programming currents. Note that the actual photon flux entering the active area was small ($\ll8.2\times10^4$), as we flood illuminated the chip with large divergence angle from $\approx \SI{2}{\centi\meter}$ distance, while the area ratio of the active area to the beam spot size was $\approx3\times10^{-10}$ (See Sec.\ref{sec:optical}). Considering that the effective bias $(I_\textrm{prog}-I_\textrm{circ})$ is the same ($~0.95I_\textrm{SW}$) at any given state (Fig.\ref{fig:fig2}b), the optical response ($\Delta I_\textrm{circ}$) should also be independent of the initial state.  Fig.\ref{fig:fig3_and_4}a shows the change in state of the device before and after the illumination, which is indeed not a function of the programming current as expected. 

To further investigate the optical response of the device, we illuminated the device using an ultrafast 
laser centered at \SI{1030}{\nano \meter}, with pulse duration $\approx\SI{400}{\femto\second}$, variable repetition rate between \SI{250}{\kilo \hertz} and \SI{4}{\mega \hertz}, and pulse energies between 10-\SI{200}{\pico \joule}. This setup allows us to control the number of pulses sent, pulse energy and pulse repetition rate. 

To demonstrate the incremental counting behavior, we have set the device to a baseline state electrically, and periodically read it every $10^5$ pulses sent from the laser. Figure \ref{fig:fig3_and_4}b shows that the device was able to accumulate up to 3 subsequent counting events at a constant programming current. It can also be observed that each count, decreasing the effective bias across the nanowire, leads to a sharp decrease in the detection efficiency as explained in the Device Operation section, which can be observed as an increased duration between successive detection events. In practical operation, following the reading of the device, the state of the cell can be reset back electrically. After reset, the detection efficiency will be restored, and the device can continue counting.

\begin{figure}[h!]
	\centering
	\includegraphics[width = 0.6\linewidth]{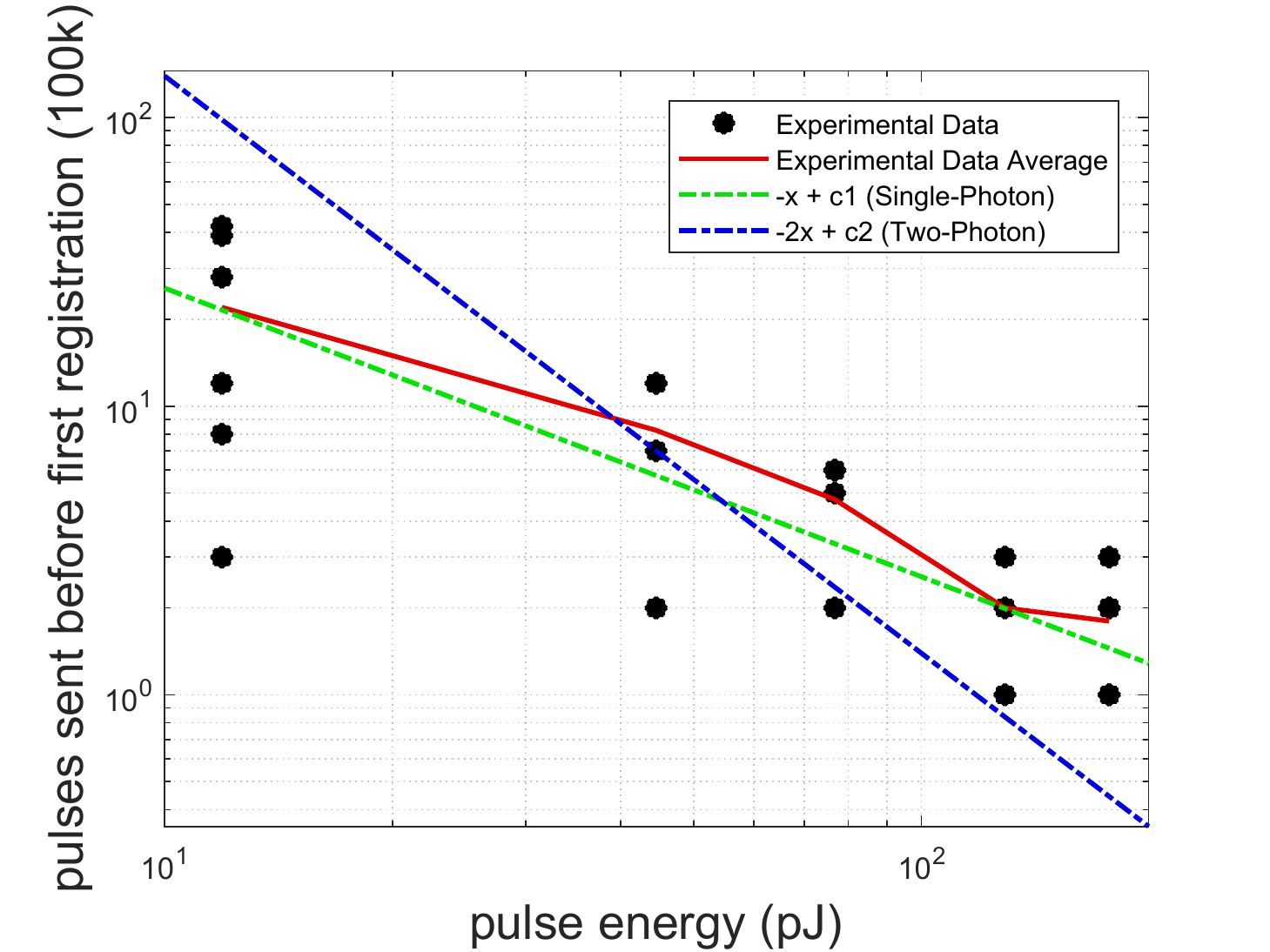}
	\caption{\textit{\textbf{Detection mechanism characterization:}}: Log-log plot for number of pulses sent before the first registration for different pulse energies at same wavelength (\SI{1030}{\nano\meter}) and pulse repetition rate (\SI{1}{\mega\hertz}). Theoretical curvatures of single-photon and two-photon mechanisms are shown as guidelines. Detection phenomena involving more than two-photons would generate theoretical curves with even larger slopes than the blue line, therefore are not shown in the graph. The optimal line slope that fits the full data is $-0.84$ with $95\%$ confidence bounds within $\pm0.31$, which includes the slope -1 as expected. The p-values estimated using a paired-sample t-test for the single-photon and two-photon hypotheses are 27.33\% and 8.21\% respectively.}
	\label{fig:fig51}
\end{figure}

So as to understand the nature of these counts (single-photon, multi-photon or bolometric), we have proceeded with changing the energy of each pulse and recorded the average number of pulses sent before the first count at a constant laser repetition rate of \SI{1}{\mega\hertz}. As shown in Figure \ref{fig:fig51}., despite the high standard deviation of the events, the energy dependence can be best explained using single-photon-dynamics. This figure is analogous to the conventional count rate vs. pulse energy graphs, with an inverted y-axis, since the number of pulses required for the first triggering event is inversely proportional to the count rate. 

Furthermore, the counting probability here is on the order of $10^{-4}\ll1$, also suggesting that the detector is operating in the single-photon regime \cite{hadfield2009single}. Note that the detector geometry ($\approx$\SI{100}{\nano\meter}$\times$\SI{300}{\nano\meter} detector area) or the material stack used in this work were not optimized for the wavelengths used in the experiments to obtain a good detection efficiency (See Sec.\ref{sec:optical}). Therefore, we suggest that it is unlikely to be detecting multi-photon events, which would imply compounded detection probabilities.

\begin{figure}[h!]
	\centering
	\includegraphics[width = 0.5\linewidth]{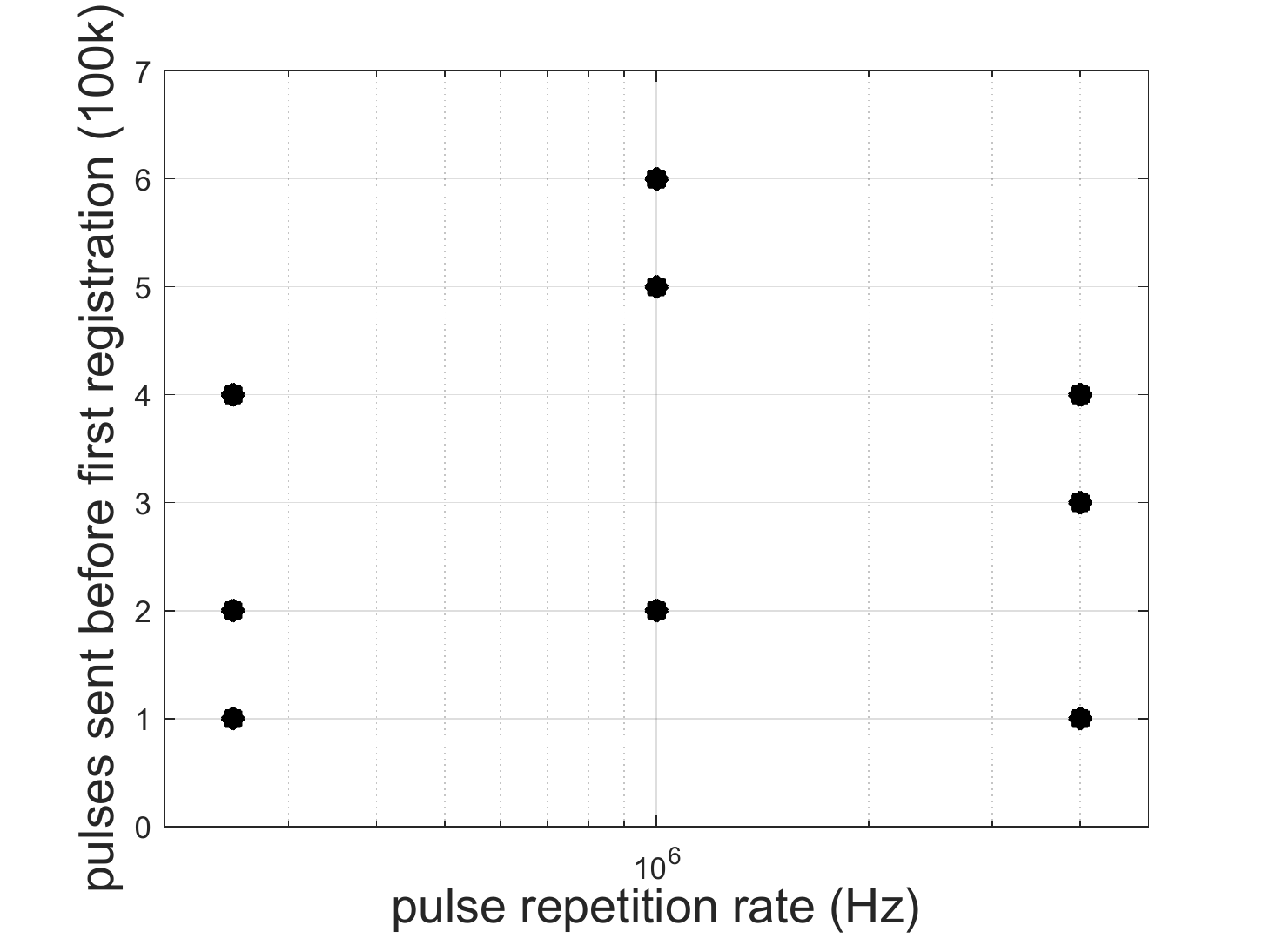}
	\caption{\textit{\textbf{Pulse repetition rate dependence experiment:}} Number of pulses sent before the first registration for different pulse repetition rates at same wavelength and pulse energy. No dependence on repetition is observed between \SI{250}{\kilo\hertz} and \SI{4}{\mega \hertz}}
	\label{fig:fig52}
\end{figure}

In order to distinguish between single-photon-based or bolometric detection mechanisms, we have sent pulses with the same energies but varying repetition rates. Doing so allowed us to controllably deliver the same number of photons with different optical powers and hence different amount of heat on the device. As shown in Fig.\ref{fig:fig52}, we have observed no correlation between photon counting probability and pulse repetition rates between \SI{250}{\kilo \hertz} and \SI{4}{\mega \hertz}.

\begin{figure}[h]
	\centering
	\includegraphics[width = 0.5\linewidth]{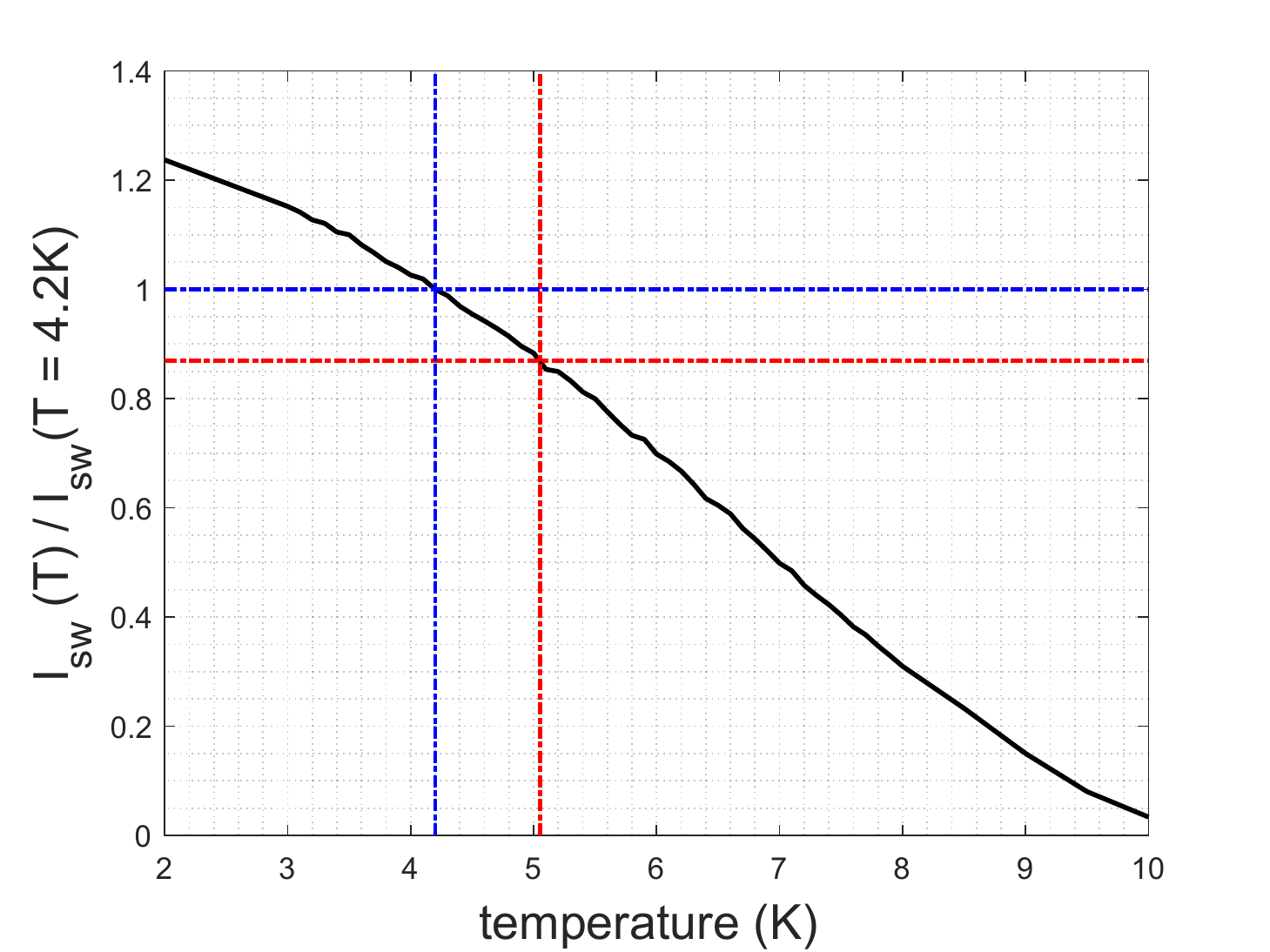}
	\caption{\textit{\textbf{Switching current of an identical nanowire versus temperature:}} Data is presented after normalization to the switching current at \SI{4.2}{\kelvin}. Blue marker indicates the liquid helium temperature (\SI{4.2}{\kelvin}) and red marker indicates $0.87I_\textrm{SW}$ (three state suppression) which occurs at \SI{5.05}{\kelvin}. The laser used in the photon-counting demonstrations has a power less than \SI{300}{\micro \watt}, which is  expected to heat the substrate no more than \SI{19}{\milli \kelvin} in immersion conditions \cite{ekin2006experimental}.}
	\label{fig:fig53}
\end{figure}

For a bolometrically operated device, the mechanism would rely on switching current suppression through the increased temperature. In order to understand if the devices worked this way, we have determined how much temperature change would be required to actuate the state changes we observed with these devices. Figure \ref{fig:fig53} shows the switching current of a stand-alone unshunted nanowire fabricated on the same chip with the same geometry, versus temperature, normalized to its switching current at \SI{4.2}{\kelvin}. 

The results show that a suppression in $I_\textrm{SW}$, that is equal to the $\Delta I_\textrm{circ}$ between 3-count-separated-states (operation limit shown in Fig.\ref{fig:fig3_and_4}b, $\Delta I_\textrm{circ} \approx0.13I_\textrm{SW}$), would require the chip temperature to increase to $\approx\SI{5.05}{\kelvin}$. Under LHe immersion and low radiation flux conditions, the  $<\SI{300}{\micro \watt \per\centi\meter\squared}$ input irradiation can increase the substrate temperature only up by \SI{19}{\milli \kelvin} which is much less than the required change for a bolometric operation \cite{ekin2006experimental}. These results strongly suggest that the process is not thermally mediated (i.e. the switching events do not occur due to the heating of the entire chip).

Finally, we have characterized the dark count performance by biasing the device for 10 hours without any illumination, while periodically performing readout. We have not observed any change in the device state for this period (See Fig.\ref{fig:sm1}).

\section{Discussion and Future Work}

In this work, we present a hybrid device that combines the advantages of SNSPDs and CCDs through single-photon to single-fluxon conversion. The device fabricated in this work can count up to three photons incrementally without any feedback circuitry. The states of these devices were nonvolatile and did not suffer from dark counts. This architecture is suitable for constructing large arrays to produce contrastive information under long-exposure and low-illumination scenarios at the expense of time-resolution of the photon arrival.  Note that the single-photon to single-fluxon conversion observed in these devices are not expected to be coherent, considering the thermal nature of the switching event. 
  
Optimization of the optical design was not pursued in this work due the complicated nature of the shunted detectors we have demonstrated. Readout of a stand-alone shunted SNSPD (i.e. not connected to a loop for post-facto measurements) would prove to be very difficult, as the low resistance change would be buried under the noise level. Moreover, as the metal layer significantly changes the behavior of the superconducting NbN (via electrical and thermal shunting as well as the proximity effect), an unshunted detector with identical geometry and superconducting film thickness cannot be used as a reference either. Therefore, a high performance readout circuitry would be necessary to further investigate and optimize the photon detection properties of these novel family of SNSPDs in separation.

On the other hand, multiple improvements can be implemented on this device idea to improve performance. First of all, the device area can be increased by replacing the short nanowire with a long, meandering geometry. This modification would also require a balance inductor on the yTron side, to keep the biasing ratio the same. Secondly, a detector with detection efficiency saturating at an earlier biasing current would increase the number of photons the device can count up to. We suggest using WSi detectors, which has shown saturated detection efficiency at bias levels down to  $0.375I_\textrm{SW}$ \cite{marsili2013detecting}. Considering that the kinetic inductivity of WSi is higher than NbN, the quantization of the states would be finer as well, which would also increase the number of states. This finer quantization might necessitate an improved design for the yTron to avoid limitations arising from readout performance. Finally, to fully implement a CCD analogous device, a method to bucket-brigade readout should be investigated. Shuttling flux between superconducting loops is an already existing technology with Josephson Junction based electronics \cite{herr2011ultra}. Implementation of such a readout scheme would further strengthen the scalability of this new device family as it makes the device count in the readout periphery much more efficient.

In conclusion, the devices demonstrated in this work have the potential to realize large-scale single-photon detectors with CCD-alike characteristics through single-photon single-flux coupling. Following the implementation of the suggested modifications, the devices can have high detection efficiency, a wider range of counting, faster readout and effective arraying as well.

\section{Acknowledgements}
Support for this work was provided in part by the DARPA Defense Sciences Office,through the DETECT program. The authors thank Andrew E. Dane, Di Zhu and John W. Simonaitis for their scientific discussions and edits to the manuscript.

\begin{singlespace}
\bibliography{main}
\bibliographystyle{plain}
\end{singlespace}

\section{Supplementary Material}
\subsection{Fabrication of the Unit Cell}
\label{sec:fabrication}
We fabricated these devices on \SI{100}{\nano\meter}  $\mathrm{Si_3N_4}$ on Si substrate. The first metal layer was patterned by a liftoff process (for the fabrication of the shunting islands). Unlike fabrication described in \cite{onen2019design}, the metal islands are defined using electron-beam lithography for the high precision definition of the shunting region. PMMA was used as the ebeam resist and was developed in 3:1 IPA:MIBK solution at \SI{0}{\degreeCelsius}. Liftoff was performed by soaking in \SI{60}{\degreeCelsius} NMP for \SI{1}{\hour}.
 
The NbN layer was then sputtered using magnetron sputtering at room temperature, following the process described in Ref.\cite{dane2017bias}. The sheet resistance of the film was measured using a four-point probe. Patterning of the superconducting layer was performed using electron beam lithography. We used ZEP520A as the resist and back-scattered electron detector for the alignment with the former layer. Exposure was done using Elionix FLS-125, at \SI{500}{\micro\coulomb/{\centi\meter\squared}} dose with \SI{500}{\pico\ampere} beam current for small features (devices) and \SI{40}{\nano \ampere} for larger features  (leads and contact pads). The resist was developed in o-Xylene at \SI{5}{\celsius} for \SI{90}{\second} followed by \SI{30}{\second} dip in isopropanol and \textrm{$N_2$} gun dry. NbN was etched with reactive ion etching using $\mathrm{CF_4}$ at \SI{50}{\watt}, for a total period of \SI{360}{\second} partitioned in 3 equal steps. Excess resist was stripped in n-methyl-2-pyrrolidone (NMP) at \SI{70}{\celsius} for \SI{1}{\hour}. Finally, the pads were defined with a direct photolithography step using a bilayer stack of PMGI-SF9 and S1813, exposed by using Heidelberg $\mu$PG 101 direct writer with \SI{7}{\milli\watt} beam at 25\% duty cycle and developed in CD-26 followed by NMP liftoff.

A Zeiss Orion SEM was used to image the devices at \SI{5}{\kilo\volt} acceleration voltage at \SI{4.8}{\milli\meter} working distance using the in-lense detector with \SI{30}{\micro\meter} aperture size.

\subsection{Electrical Characterization}
\label{sec:electrical}
A plurality of devices were fabricated in this study, using different material stacks and geometries. We have observed that the devices that could be programmed using electrical signals, could also be programmed optically with varying sensitivities and noise levels. The results reported in this work are based on one hero device out of these trials. 

In Fig.\ref{fig:fig2}b, the device was first set to an arbitrary state electrically. Consequently, $I_\textrm{prog}$ was incrementally increased and the first programming voltage that modifies the state is recorded. At this point, the total current flowing through the nanowire $(I_\textrm{prog}-I_\textrm{circ}) = I_\textrm{SW}$. Similarly, when the same procedure was conducted by decreasing the programming current, the same equation holds for $-I_\textrm{SW}$ when the state decrements. Therefore, both currents in terms of $I_\textrm{SW}$ can be calculated at any given point on this curve. Once the $I_\textrm{circ}$ was found in relation to the $I_\textrm{SW}$, with the additional information of the switching current measured on a sister-device (\SI{62}{\micro\ampere}) allowed us to find the separation between these states are worth of \SI{2.7}{\micro \ampere}.

Note that due to the imperfections of the yTron, most of the negative states (counterclockwise currents) resolved in the same output, leading to utilization of 25 states out of 46. We have determined this property using the sign of the $I_\textrm{circ}$ for different states, where ideally, there should be an equal number of states on both polarities.

\subsection{Optical Characterization}
\label{sec:optical}
The results shown in Fig.\ref{fig:fig3_and_4}b are conducted with an ultrafast laser centered at \SI{1030}{\nano\meter}, with a repetition rate of \SI{1}{\mega \hertz} and pulse energy $\approx$ \SI{80}{\pico\joule}, which means every pulse consists of $\approx$ $4\times10^8$ photons. The beam spot size of this laser was $\approx$\SI{1}{\centi\meter\squared}, which gives an area ratio (between the spot size and detector) of $3\times10^{-10}$. The first detection event is registered after the first readout which is performed after sending $10^5$ pulses, which is equivalent to $\approx$ $1.2\times10^4$ thousand photons (i.e., a counting probability of $~8.3\times10^{-5}$.). 

\begin{figure}[h!]
	\centering
	\includegraphics[width = 0.5\linewidth]{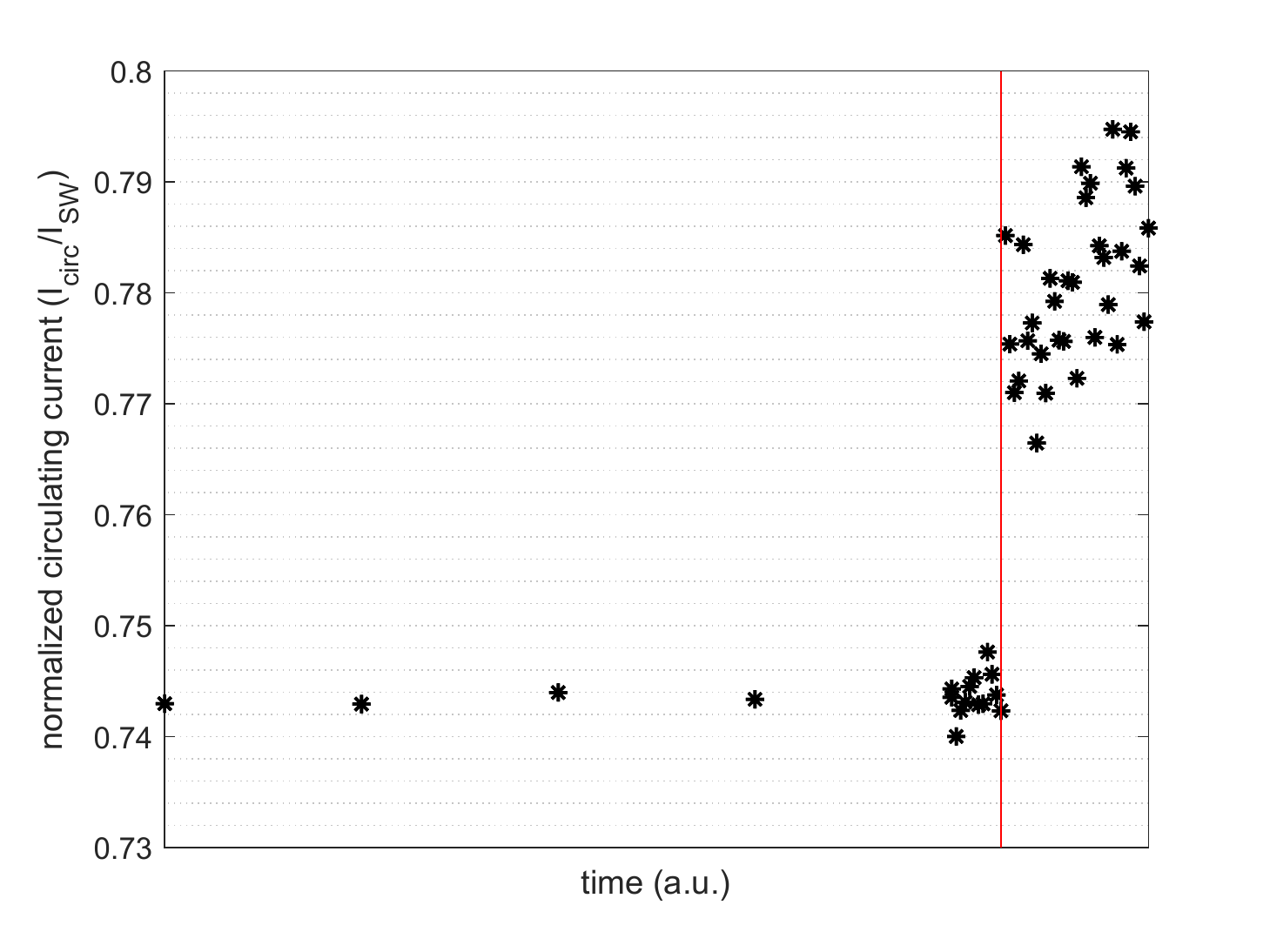}
	\caption{\textit{\textbf{Dark count experiment:}}: The device was biased and state was monitored for \SI{10}{\hour} with \SI{2.5}{\hour} intervals. No state change was observed during this period. Following this period, the device was illuminated to check operation, indicated by the red line.}
	\label{fig:sm1}
\end{figure}

The dark count rate is characterized by biasing the device for 10 hours without any illumination, while periodically performing readout. We have not observed any change in the device state for this period (Fig.\ref{fig:sm1}).

\end{document}